\documentclass[12pt,preprint]{aastex}

\shorttitle{Angular Differential Imaging}

\shortauthors{Marois et al.}

\begin{document}

\title{Angular Differential Imaging: a Powerful High-Contrast Imaging Technique \footnote{Based on observations obtained at the Gemini Observatory, which is operated by the Association of Universities for Research in Astronomy, Inc., under a cooperative agreement with the NSF on behalf of the Gemini partnership: the National Science Foundation (United States), the Particle Physics and Astronomy Research Council (United Kingdom), the National Research Council (Canada), CONICYT (Chile), the Australian Research Council (Australia), CNPq (Brazil) and CONICET (Argentina).}}

\author{Christian Marois$^{2,3}$, David Lafreni\`{e}re$^{2}$, Ren\'{e} Doyon$^{2}$, Bruce Macintosh$^3$,\\ Daniel Nadeau$^{2}$}
\affil{$^{2}$ D\'{e}partement de physique and Observatoire du Mont M\'{e}gantic, Universit\'{e} de Montr\'{e}al, C.P. 6128, Succ. A,\\ Montr\'{e}al, QC, Canada H3C 3J7}
\affil{$^{3}$ Institute of Geophysics and Planetary Physics L-413,\\ Lawrence Livermore National Laboratory, 7000 East Ave, Livermore, CA 94550}
\email{cmarois@igpp.ucllnl.org david@astro.umontreal.ca doyon@astro.umontreal.ca bmac@igpp.ucllnl.org nadeau@astro.umontreal.ca} 

\begin{abstract} 
Angular differential imaging is a high-contrast imaging technique that reduces
quasi-static speckle noise and facilitates the detection of
nearby companions. A sequence of images is acquired with an
altitude/azimuth telescope while the instrument field derotator is switched
off. This keeps the instrument and telescope optics aligned and allows the field of view to rotate with respect to the instrument. For each image, a reference PSF is constructed from other appropriately-selected images of the same sequence and subtracted to remove quasi-static PSF structure.  All
residual images are then rotated to align the field and are
combined. Observed performances are reported for Gemini North data.
It is shown that quasi-static PSF noise can be reduced by a factor $\sim $5 for each image
subtraction. The combination of all residuals then provides an
additional gain of the order of the square root of the total number
of acquired images. A total speckle noise attenuation of 20-50 is obtained for one-hour long observing sequences compared to a single 30s exposure. A PSF noise attenuation of 100 was achieved for two-hour long sequences of images of Vega, reaching a 5-sigma contrast of 20 magnitudes for
separations greater than 8$^{\prime \prime}$. For a 30-minute long sequence, ADI achieves 30 times better signal-to-noise than a classical observation technique. The ADI technique can be used with currently available instruments to search for $\sim 1$ M$_{\rm{Jup}}$ exoplanets with orbits of radii between 50 and 300~AU around nearby young stars. The possibility of combining the technique with other high-contrast imaging methods is briefly discussed.
\end{abstract} 
\noindent{\em Suggested running page header:} Angular Differential Imaging

\section{Introduction}

Direct detections of very faint exoplanets and brown dwarfs near bright stars are essential to understand substellar formation and evolution 
around stars. This endeavor is now one of the major goals for next
generation 10-m telescope instruments and future 30- to 100-m
telescopes. The task is dauntingly difficult. The exoplanet or brown
dwarf image is usually much fainter than the background from the
brilliant stellar point spread function (PSF) image. Besides the Poisson
noise, ground-based telescopes suffer from atmospheric turbulence
that produces random short-lived speckles that mask faint companions. If
these two limitations were the only ones, a simple solution would be to
integrate longer to average these random noises and gain as the square root of the integration time. But observations have shown that, for integrations longer than a few minutes, the PSF noise converges to a quasi-static noise
pattern, thus preventing a gain with increasing integration time
\citep{marois2003,marois2005,masciadri2005}. To achieve better detection limits, it is
thus necessary to subtract the quasi-static noise using a reference
PSF. Both ground- and space-based imaging are plagued with this stellar
PSF calibration problem caused by imperfect optics and slowly evolving optical alignments. For ground-based imaging, subtraction of a reference PSF obtained from a star close to the target achieves a factor $\sim 4$ of PSF noise attenuation, leaving residuals that are also quasi-static and thus severely limiting detection of fainter companions \citep{marois2005}. For space telescopes that have a better PSF stability, like HST, a partial solution was found by subtracting two stellar images acquired during the same orbit with a different roll angle. This technique, called ``roll deconvolution'', successfully subtracts the stellar image by a factor 50 \citep{schneider2003,fraquelli2004} but is also ultimately limited by PSF evolution. A similar technique, called angular differential imaging (ADI), can be used on ground-based altitude/azimuth telescopes to subtract a significant fraction of the stellar quasi-static noise and can potentially achieve detection limits that improve as the square root of the integration time.

In this paper, the ADI technique is described and its performance is analyzed using a simple analytical model and data from the Gemini Altair adaptive optics system and the NIRI near infrared camera. The PSF stability with Altair/NIRI is studied and its impact on
ADI performances is discussed. Detection limits for three stars of our
ongoing young nearby star survey are then shown. A comparison between
ADI and normal imaging is also presented. Finally, the
possibility of using ADI with other high-contrast imaging techniques is
discussed. 

\section{The Angular Differential Imaging Technique \label{aditech}}
ADI is a PSF calibration technique that can, in principle, suppress the
PSF quasi-static structure by a large factor
\citep{marois2004phd}. It consists of the acquisition of a sequence of
images with an altitude/azimuth telescope and the instrument rotator
turned off (at the Cassegrain focus) or adjusted (Nasmyth) to keep the instrument and telescope optics aligned. This setup improves the stability of the quasi-static PSF
structure throughout the sequence while it causes a slow rotation of the
field of view (FOV) with respect to the instrument. Note that only the FOV, not the PSF, rotates with time. Since the FOV rotates during an exposure, companion PSFs are smeared azimuthally. This effect increases linearly with angular separation and slightly decreases companions peak intensity. Short exposures and the use of an optimized aperture photometry box can minimize this effect. For each image, after data reduction and image registration of the whole sequence, a reference PSF obtained from other
images of the same sequence is subtracted to remove the quasi-static
structure. Given enough FOV rotation during the sequence, this subtraction preserves the signal from any eventual companion. All the image differences are then rotated to align the FOV and are median combined. 

This technique offers a number of advantages over more classical
ground-based observations since the target observations themselves are
used to construct a reference PSF. This means that the reference PSF has the same spectrum and brightness as the target and that no time is lost to acquire reference observations of
a different target. Ghost images from optical reflections and the sky
flux are also removed by the subtraction. The detector flat field errors are
averaged since the FOV is integrated with different pixels as it rotates on the detector. The ADI technique is a generalization of the
``roll deconvolution'' technique used with HST since
several images, each at a different field angle, are acquired and
combined. A technique similar to ADI has been developed independently
by \citet{liu2004} to search for circumstellar disks.

\section{Noise Attenuation Theory with ADI\label{optpess}}

For each reduced and registered image of an ADI sequence, a reference PSF has to be built from the
images of the same sequence. The way that this reference PSF is
built is of great importance since it directly affects the noise
attenuation performance. We have used two methods to construct the
reference PSF.

The first method is simply to take the median of all the images of the
sequence. If enough field rotation has occurred during the sequence so that an
eventual point source has moved by at least twice its full width at half maximum (FWHM), then this point source will be largely rejected by the median which will leave only the average PSF. The minimum radial separation at which this occurs is noted $R_{\rm{min}}$. Since the median is taken
over a large number of images, the pixel-to-pixel noise (i.e. PSF, flat field, dark and sky Poisson noises and detector readout noise) of the reference image is much less than that of any individual image. Thus this first method
minimizes the noise in regions where the residuals are limited by
pixel-to-pixel noise. However, since a sequence typically lasts more than an hour, the reference PSF only has modest quasi-static speckle correlation with the individual images of the sequence.

The second method is to take the median of only a few images as
close in time as possible but for which the displacement due to
field rotation between the images is, at a given separation, at least 1.5 PSF FWHM. This
displacement ensures that the flux inside the PSF core of an
eventual point source is not significantly reduced by the subtraction. The time $\tau_{\rm{min}}$ required for such field
rotation is function of the separation angle from the target, the target azimuth $A$ and zenith distance $z$ and the telescope latitude $\phi$. The rotation rate $\psi$ (degree/minute) of the FOV is
obtained from the time derivative of the parallactic angle and is given \citep{mclean1997} by

\begin{equation}
\psi = 0.2506 \frac{\cos A \cos \phi}{\sin z}
\end{equation}

\noindent Figure 1 is
provided as a reference to determine the time interval for observations from the summit of the Mauna Kea. 
This second technique provides better quasi-static speckle noise
attenuation since the reference PSF is built using images acquired
at short time intervals. However, the pixel-to-pixel noise of the
reference image may not be negligible compared to that of an individual
image.

The first method is optimized for regions where the residuals are
limited by pixel-to-pixel noise while the second is optimized for
regions where the residuals are still limited by speckle noise.
The combination of both techniques into a single reduction algorithm
will be discussed in section~\ref{adialgo}.

The ADI technique attenuates the PSF noise in two steps: (i) by subtraction of
a reference image to remove correlated speckles and (ii) by the combination of all residual images after FOV alignment to average the residual noise. The noise attenuation $N/\Delta N$ is defined as the ratio of the local noise $N$ in an image over the noise $\Delta N$ of the residual image.

The noise attenuation obtained by the subtraction of the reference
PSF, $\left[ N/\Delta N \right]_{\rm{S}}(\theta, \tau, t_{\rm{exp}})$, is a function of the angular separation, $\theta$, the time interval $\tau$ between the image and its reference and the individual image exposure time $t_{\rm{exp}}$ (including overheads). Strong quasi-static speckle correlation between successive images leads to strong attenuation and thus better detection limits for a given total integration time. A good and stable seeing is thus expected to deliver better ADI quasi-static speckle attenuation. For a perfect case when all static and quasi-static speckles have been removed by ADI, detection limits are ultimately limited by short-lived atmospheric speckles that have a correlation timescale of a few tens of ms, shorter than the time interval required to obtain sufficient FOV rotation to build the ADI reference PSF.

The additional noise attenuation resulting from the combination of $n$ de-rotated residual images is function of the correlation of those images. This attenuation is given by $\sqrt{n_{\rm{eff}}}$, where $n_{\rm{eff}}$ is the effective number of uncorrelated residual images; if all residuals are uncorrelated, then $n_{\rm{eff}}=n$. Equivalently, for an image sequence of total time $t = t_{\rm{exp}}n$, one has to wait for a time

\begin{equation}
\tau_{\rm{dcorr}} = \frac{n}{n_{\rm eff}} t_{\rm exp}\label{eq-1}
\end{equation}

\noindent before two de-rotated residual images are decorrelated. Thus the total noise attenuation $N/\Delta N$ will be

\begin{equation}
\frac{N}{\Delta N}(t,\theta, \tau, t_{\rm{exp}}) \cong \sqrt{\frac{t}{\tau_{\rm{dcorr}}}} \left[ \frac{N}{\Delta N}\right]_{\rm{S}} (\theta, \tau, t_{\rm{exp}}) \rm{.}\label{eq0}
\end{equation}

For simplicity, let us consider a dominant speckle noise having a correlation timescale $\tau_{\rm{speck}}$. The ADI noise attenuation resulting from the combination of all difference images is function of $\tau_{\rm{speck}}$. Its behavior can be defined for two limiting regimes: a) when either $\tau_{\rm{speck}}$ is much longer than $\tau$ or shorter than $t_{\rm{exp}}$ and b) $\tau_{\rm{speck}}$ is longer than $t_{\rm{exp}}$ but shorter than $\tau$.

In the first regime, the residuals of consecutive image differences are decorrelated, either because
the correlated structure of the PSF which lasts for long times ($\tau_{\rm{speck}} \gg \tau$) has been
removed, leaving only Poisson noise ($\tau_{\rm{dcorr}} = t_{\rm{exp}}$), or because the PSF
structure was already uncorrelated between consecutive images in the
first place ($\tau_{\rm{speck}} \le  t_{\rm{exp}} \Rightarrow \tau_{\rm{dcorr}} = t_{\rm{exp}}$). In this case, the
speckle attenuation increases with the square root of the total number
of image differences. The decorrelation timescale is simply

\begin{equation}
\tau_{\rm{dcorr}} = t_{\rm{exp}}\rm{.}\label{eq1}
\end{equation}

In the second regime, since $\tau_{\rm{speck}}$ is longer than $t_{\rm{exp}}$ but shorter than $\tau$, the residuals of consecutive
differences are partially correlated and do not average-out as efficiently. Assuming that only the noise having a spatial
scale $\sim 1$ PSF FWHM limits point source detection, the time needed to decorrelate this noise is the shortest time between $\tau_{\rm{FWHM}}$, the time for a one FWHM FOV rotation, and $\tau_{\rm{speck}}$. The decorrelation timescale is thus

\begin{equation}
\tau_{\rm{dcorr}} \cong \rm{MIN} (\tau_{\rm{FWHM}},\tau_{\rm{speck}})\rm{.}\label{eq2}
\end{equation}

We note that if $\tau$ is several minutes and if individual ADI differences are dominated by random short-lived speckles, then the noise attenuation can follow Eq.~\ref{eq1} for several images but ultimately converges to Eq.~\ref{eq2} when sufficient averaging of short-lived speckles has occurred.

It is important to known that reducing $t_{\rm{exp}}$ to increase the number of images for a given image sequence of length $t$ would not result in better detection limits. For the first regime, shorter $t_{\rm{exp}}$ would mean a smaller companion S/N per image; the final S/N would thus be the same (neglecting losses due to overheads). For the second regime, a shorter $t_{\rm{exp}}$ would yeild quasi-static speckles correlated over more consecutive images, also resulting in the same performance.

We emphasize that ADI guarantees a gain in detection with increasing observing time for both regimes. In the worst case, the speckle attenuation efficiency is limited by field rotation. This is an enormous advantage over classical observation in which the quasi-static aberrations prevent significant gain after a relatively short observing time. An equation similar to Eq.~\ref{eq2} also applies to classical observations, but in this case there is no $\tau_{\rm{FWHM}}$ since the field is not rotating. Speckle attenuation is then governed by $\tau_{\rm{speck}}$, which can be very large, reducing drastically the efficiency of the observations. Section~\ref{adivscla} presents a comparison of ADI and classical observations.

\section{Observations}

The ADI technique was first used at the Gemini North telescope using the Altair adaptive
optic system \citep{herriot1998} and the near-infrared camera NIRI \citep{hodapp2000} in queue
mode. Observations and detailed results obtained for three stars of our nearby
young star survey, Vega, HD18803 and HD97334B are presented. Data for two other stars, HIP18859 and HD1405 will also be discussed for comparison since they have been acquired with a different technique. The exposure time was fixed to 30s, a time short enough to minimize PSF saturation and the off-axis PSF smearing due to the FOV rotation during the exposure and long enough to get a good observing efficiency and to be limited by the sky background rather than by the read noise at wide separations.

For Vega, HD18803 and HD97334B, the imaging sequence consisted in the acquisition
of a series of saturated images in the off methane $1.58$~$\mu$m 6.5\%
bandwidth filter with the cassegrain rotator turned off. This filter was chosen to minimize the brightness ratio of the star to that of
methanated exoplanets or brown dwarfs. Unsaturated short exposures were
acquired before and/or after the saturated sequence to calibrate detection
limits.

For Vega, data were obtained on 2004 August 26 and 2004 September 1 (program GN-2004A-Q-11). In total, 225 and 177 30s
exposures were respectively acquired on the 26th and September 1st. Because of Vega's
brightness, short exposure PSFs of a nearby reference star were acquired
for photometric calibration. Seeing conditions were excellent on the 26th and average on the 1st. Images were saturated inside a $\sim$6$^{\prime \prime}$ diameter. For
HD18803, 90 30s exposures were obtained on 2004 December 24 (program GN-2004B-Q4) and, for
HD97334B, 90 30s exposures were acquired on 2005 April 18 (program GN-2005A-Q16). For 
HD18803 and HD97334B, seeing conditions were average to good and PSFs were
saturated inside diameters of 0.7$^{\prime \prime}$. Table~\ref{tabobs} summarizes the observations. Strehl ratios were obtained by analyzing unsaturated data acquired before and after each sequence and by comparing the PSF peak intensity with that of a simulated unaberrated PSF having the same pixel sampling, bandpass and integrated flux. It can be
deduced from table~\ref{tabobs} that since $R_{\rm{min}}$ is less than
the saturated radius for all three targets, ADI can be applied at all
separations to detect point sources.

As part of Altair science verification we have obtained a sequence of
observations of the star HIP18859 on 2003 November 18 (program GN-2003B-SV-102). During this
sequence, the filter was switched from the broadband $H$ filter to a
narrow band filter every fourth exposure to acquire unsaturated
images. This data set will be discussed in section~\ref{psfevol}. 

Observations of the star HD1405 were obtained on 2004 August 23 (program GN-2004B-Q-14)
with the instrument rotator operating to keep the FOV orientation
fixed throughout the sequence. In total, 38 30s exposures were
obtained. These observations will be used in section~\ref{adivscla} for
a comparison between ADI and classical observations.

\section{ADI Data Reduction Algorithm}

\subsection{Preliminary Data Reduction}\label{reduc}

The data reduction consists of flat field normalization, bad pixel
correction using a median over surrounding pixels, and distortion
correction using software provided by the Gemini Staff (Trujillo, private
communication) and modified to use the IDL
{\sl interpolate} function with cubic interpolation. Images were then
copied into larger blank images to ensure that no FOV was lost when
shifting and rotating images. The center of the PSF of the first image
of the sequence was then registered to the image center by minimizing
the diffraction spikes residuals after subtraction of a 180-degree rotation 
of the image. The rest of the images were then registered by
cross-correlation of the diffraction spikes with the first image. An azimuthally symmetric intensity profile was finally
subtracted from each image to remove the smooth seeing halo.

\subsection{ADI algorithm}\label{adialgo}

As discussed in section~\ref{optpess}, two methods can be used to subtract the
quasi-static PSF structure: subtracting the median of all images or
subtracting a reference PSF obtained from a few images acquired
as close in time as possible. These two methods can be combined into a
single algorithm that optimizes speckle subtraction and minimizes
pixel-to-pixel noise. First, the median of all the images is subtracted from
each individual image. For a sequence 

\begin{equation}
I_1(t_1,\theta_1), I_2(t_2,\theta_2), I_3(t_3,\theta_3), \ldots, I_n(t_n,\theta_n)
\end{equation}

\noindent of $n$ reduced and registered images (see section~\ref{reduc}), where $t_i$ is the mean time of exposure $i$ and $\theta_i$ is the FOV orientation at time $t_i$, the first reference subtraction is simply

\begin{equation}
I_i^D = I_i - \rm{median}(I_1, I_2, I_3, \ldots, I_n)
\end{equation}

\noindent An optimized reference PSF (second method) is then obtained for each image by median combining 4 images (two acquired before and two after) that show at least a 1.5 FWHM (one FWHM in the off methane 6.5\% filter is equal to $\sim 3$~pixels) FOV orientation difference. This choice insures that the average $\tau$ of the reference PSF is $\sim $0 and that the linear time-evolution of quasi-static speckles having $\tau_{\rm{speck}} \ge 2\tau$ is removed. For the construction of this reference PSF,
the image is broken into many annuli to accommodate for the dependence of
$\tau_{\rm{min}}$ on the separation. The intensity of the reference PSF
is then scaled appropriately inside each annulus to minimize the noise after subtraction. The scaling factor converges to zero if the
annulus is dominated by pixel-to-pixel noise or to unity if it is
dominated by correlated speckles. The optimized reference PSF is then
subtracted. This step can be summarized by the following equation for each annulus

\begin{equation}
I_i^{ADI} = I_i^D - a \times \rm{median}(I_{i-b}^D,I_{i-b-1}^D,I_{i+c}^D,I_{i+c+1}^D)
\end{equation}

\noindent where $a$ is a normalizing factor to minimize the noise inside the annulus and $b$ and $c$ are the number of images required to get at least a 1.5 FHWM FOV orientation difference between $I_i$ and the images acquired respectively before and after $I_i$. Differences are then rotated to align the FOV to that of the
first image. Finally, a median is taken over all
differences. The final combination step is thus

\begin{equation}
I^{ADI}_F = \rm{median}\left[ I_1^{ADI},rot(I_2^{ADI},\Delta\theta_{1-2}), rot(I_3^{ADI},\Delta\theta_{1-3}), \ldots, rot(I_n^{ADI},\Delta\theta_{1-n})\right]
\end{equation}

\noindent Optionally, the final residual image may be convolved by a
Gaussian of FWHM equal to that of the PSF to attenuate further the high
frequency noise. Table~\ref{tabadi} summarizes the entire ADI reduction
algorithm. 

\section{Results}

\subsection{PSF Evolution Time-Scale}\label{psfevol}

The PSF noise evolution timescale can be studied through the evolution of the
noise attenuation $\left[ N/\Delta N \right]$ for the difference of two images as a function of the time interval, $\tau$. The noise attenuation is the ratio of $N$ and $\Delta N$, which are the measured rms noises in an annulus of width equal to one PSF FWHM in the original and in the difference images respectively. For this analysis, all images have first been unsharp masked using an $8\times 8$ FWHM box to remove low-frequency noise and then median filtered with a $1\times 1$ FWHM box to remove hot/bad pixels. This step is necessary to prevent biasing the noise estimate $N$ of single images and leaves only speckles that have a spatial scale of the order of one FWHM. Images were subtracted two by two with increasing time interval. Fig.~\ref{fig2} shows the noise attenuation for Vega, HD18803 and HD97334B for angular separations of 2, 4, 6 and 8$^{\prime \prime}$.

\begin{figure}
\epsscale{1}
\plotone{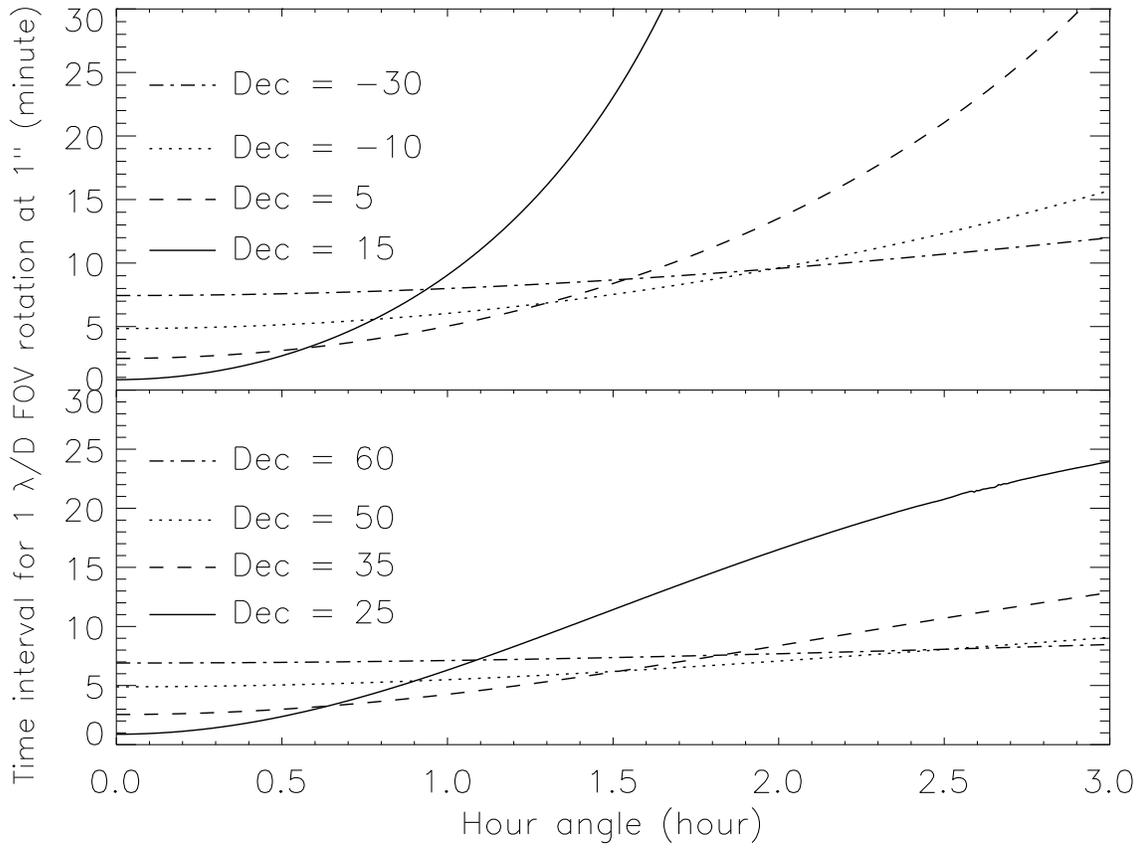}
\caption{Interval of time required for a point source to move by 1
  $\lambda/D$ (1.6~$\mu $m on a 8-m diameter telescope) at 1$^{\prime \prime}$ as a function of hour angle for various declinations. Calculated for Mauna Kea, Hawaii.\label{fig1}}
\end{figure}

All noise attenuation curves show stronger noise attenuation for shorter time intervals. For HD18803 and HD97334B, the noise attenuation reaches $\sim 2-5$ inside 4$^{\prime \prime}$ for $\tau \sim$2 minutes. At larger separation, the residuals are limited by pixel-to-pixel noise. For separations less than 4$^{\prime \prime}$, the noise attenuation drops by a factor 2 after approximately 60 minutes. For Vega, the noise attenuation reaches $\sim 4-6$ at
all separations for $\tau \sim 1$ minutes and drops by a factor 2 after
approximately 15 to 20 minutes. The stronger noise attenuation achieved on Vega and HD97334B for short time intervals could be explained by better seeing that stabilizes the structure and enables a better subtraction. Additionally, since $\tau_{\rm{min}}$ is less than 10 minutes at 1$^{\prime \prime}$ within $\pm 1$h from the meridian for all targets having a declination $-30^{\circ}$ to $+65^{\circ}$ at Mauna Kea (see Fig.~\ref{fig1}), ADI can be used with little loss of speckle attenuation ($\sim $30\%) on most of the available sky for separations greater than $\sim 0.5^{\prime \prime}$.

Analysis of the HIP18859 data set that included a frequent filter change to acquire unsaturated images shows that a drop by a factor of 2 in speckle attenuation occurs following each filter change. This evolution of the PSF structure is probably due to the filter wheel not returning to its exact position after each change. This suggests that the best observing procedure is to prevent any alteration of the optical setup during an observing sequence to maximize the PSF noise stability.

\subsection{ADI Speckle Attenuation}

\begin{figure}
\epsscale{1}
\plotone{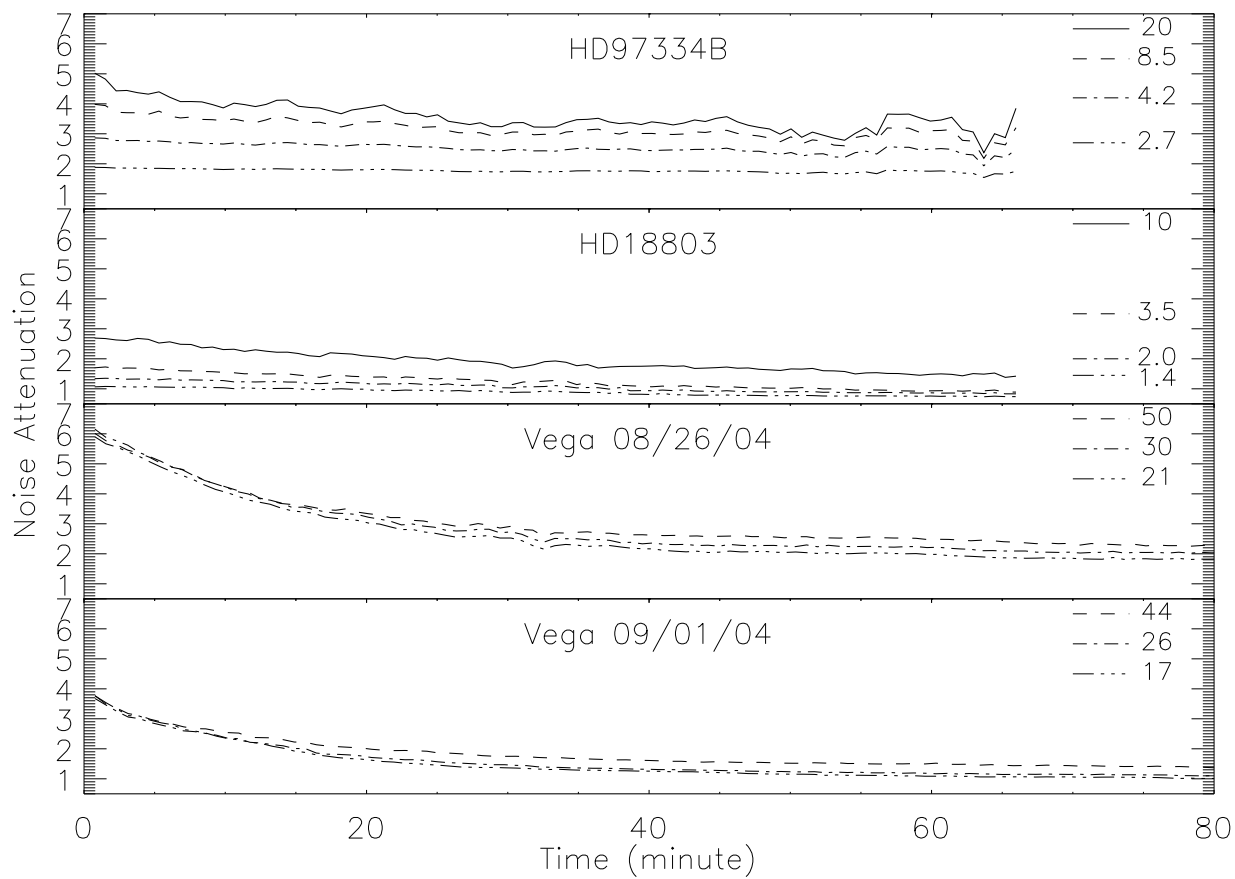}
\caption{Noise attenuation obtained by subtracting images two by two with increasing time interval for HD97334B, HD18803 and Vega acquired on the 08/26/04 and on the 09/01/04. Solid, dashed, dot-dashed and triple dot-dashed lines are for 2$^{\prime \prime}$, 4$^{\prime \prime}$, 6$^{\prime \prime}$ and 8$^{\prime \prime}$ respectively. For Vega, there is no solid line for 2$^{\prime \prime}$ since images are saturated at that separation. Small lines show the estimated noise attenuation limit imposed by photon, sky, flat, read and dark noises.\label{fig2}}
\end{figure}

\begin{figure}
\epsscale{1}
\plotone{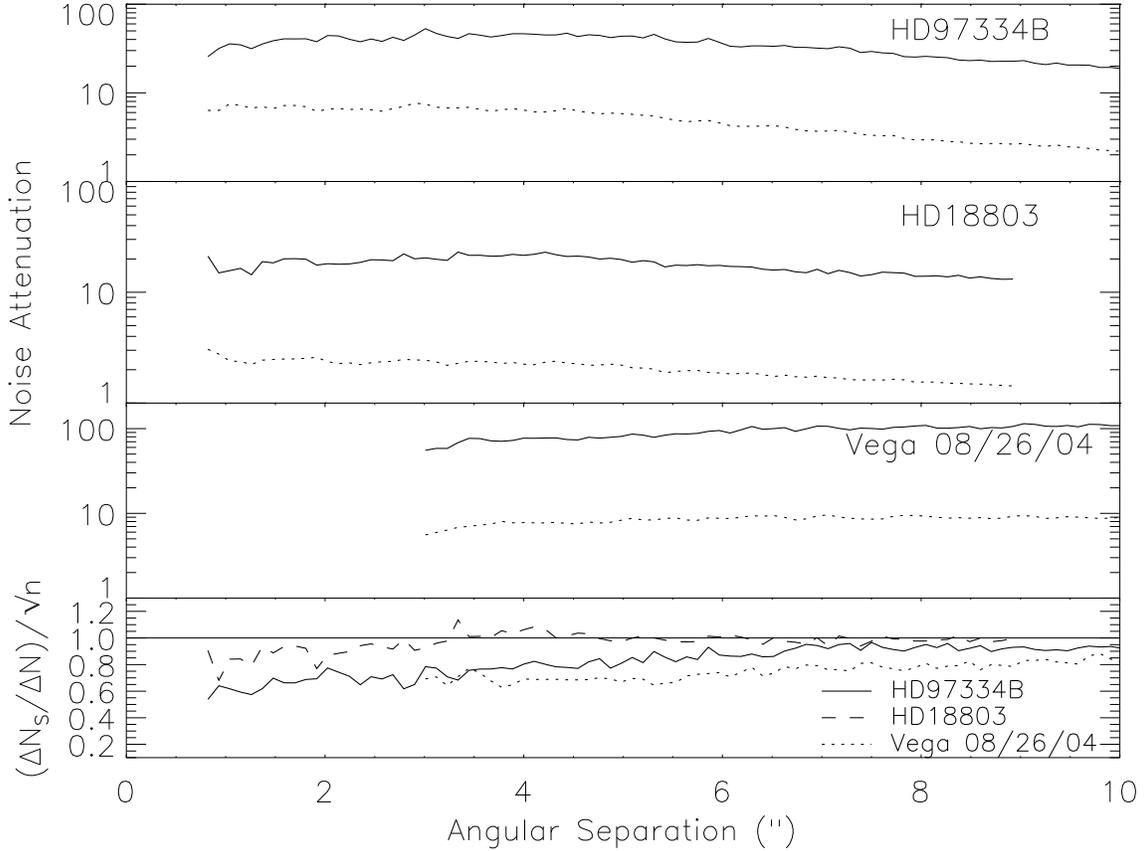}
\caption{Single ADI difference $\left[ N/\Delta N \right]_{\rm{s}}$ and total ADI $\left[ N/\Delta N \right]$ noise attenuations with separation for HD97334B, HD18803 and Vega. Dotted lines show attenuation from a single ADI difference while solid lines show the attenuation after median combining all ADI differences. The bottom panel shows the ratio of the speckle noise attenuation of the combined ADI difference $[N/\Delta N]$ over that of a single ADI difference $[N/\Delta N]_{\rm{S}}$ normalized by the square root of the total number $n$ of images in each sequence. Data for HD18803 are truncated at 9$^{\prime \prime}$ due to a PSF decenter that brought a part of the field $> 9^{\prime \prime}$ outside the FOV.\label{fig3}}
\end{figure}

Fig~\ref{fig3} shows for all ADI targets the average noise attenuations $[N/\Delta N]_{\rm{S}}$ achieved for one image difference and the noise attenuation $[N/\Delta N]$ obtained after median combining all the image differences. Again, an $8 \times 8$ FWHM unsharp mask and a $1 \times 1$ FWHM median filter were applied to each image to produce this figure. Each ADI difference attenuates by a factor 2-10 the quasi-static speckle noise. Note that these attenuations are better than what is observed in Fig.~\ref{fig2} since here the optimized reference PSF is the result of a median of 4 images, 2 acquired before and 2 after. Random speckles are averaged and the linear speckle evolution with time of speckles having at least $\tau_{\rm{speck}} \ge 2\tau_{\rm{min}}$ is removed. Similarly to what was observed in Fig~\ref{fig2}, better seeing conditions (HD97334B compared to HD18803) seem to be related to better speckle noise attenuation. A total noise attenuation of $\sim 35$ between 0.8 and 10$^{\prime \prime}$ was obtained for HD18803 and HD97334B, while this attenuation
reached 100 for Vega (August 26th data). The higher attenuation for Vega comes partly from
the larger number of images (225 {\sl vs} 90) and partly from better seeing
conditions which provided a better attenuation from single image
subtractions. HD97334B and HD18803 noise attenuations are shown to improve at all separations down to the detector saturation limit and clearly below it if we extrapolate the performances shown at 0.8$^{\prime \prime}$.

The bottom panel of Fig.~\ref{fig3} is generated by dividing the ratio of the noise attenuation of the combined ADI difference $[N/\Delta N]$ over the noise of a single ADI differences $[N/\Delta N]_{\rm{S}}$ by the expected noise attenuation $\sqrt{n}$ if the noise was decorrelated. From Eq.~\ref{eq0}

\begin{equation}
R = \sqrt{\frac{t_{\rm{exp}}}{\tau_{\rm{dcorr}}}}\rm{.}\label{eqR}
\end{equation}

Because the pixels affected by the secondary mirror support diffraction spikes are masked from the images, the effective number of images combined is reduced at some separations. The bottom panel of Fig.~\ref{fig3} has been corrected for this effect, which can reach 20\% at small separations. For separations greater than 2$^{\prime \prime}$, all three targets achieve more than 70\% of the $\sqrt{n}$ attenuation expected if residuals are decorrelated. At smaller separations the attenuation is lower, revealing a correlated noise in successive ADI differences that lowers the number of independent images (see Eq.~\ref{eq2}).

\begin{figure}
\epsscale{1}
\plotone{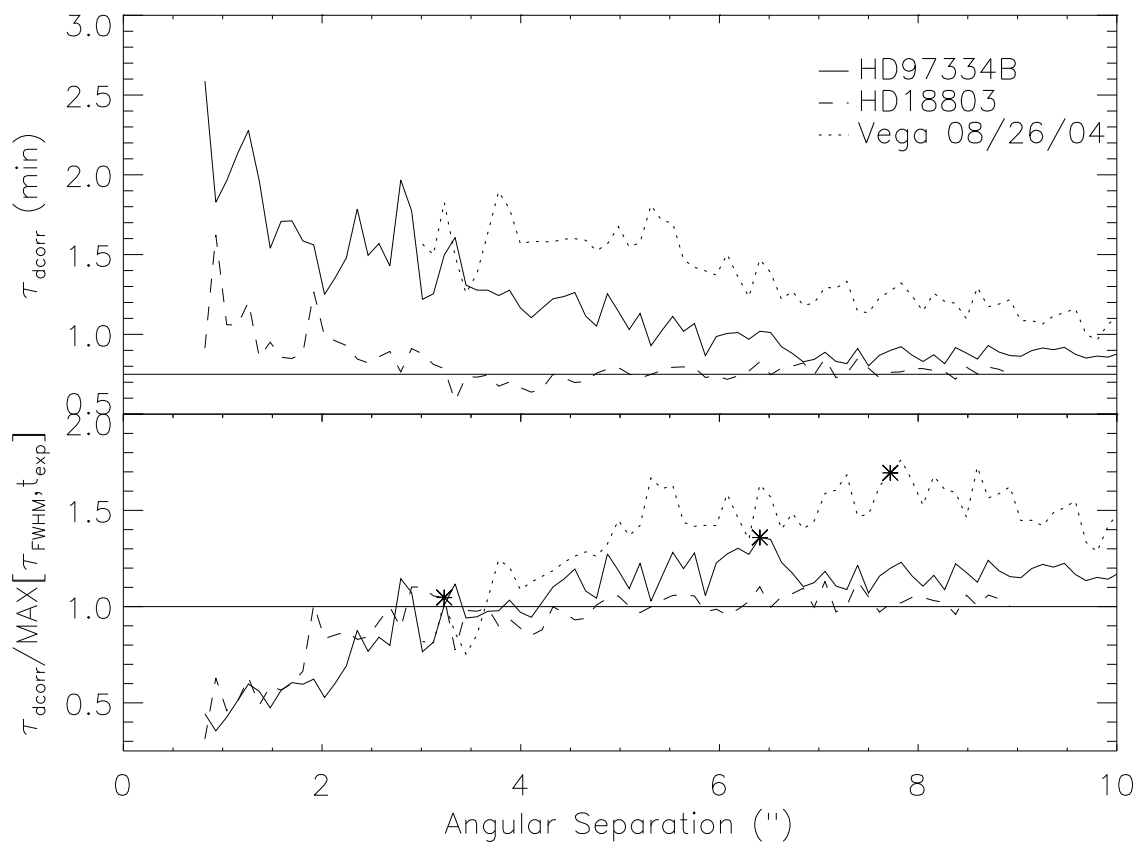}
\caption{Upper panel: Estimated residual speckle decorrelation timescale $\tau_{\rm{dcorr}}$ as a function of angular separation for HD97334B, HD18803 and Vega (August 26th). Bottom panel: Normalized residual speckle decorrelation time as a function of angular separation for the three same targets. The star symbol on each curve indicates at what separation $\tau_{\rm{FWHM}} = t_{\rm{exp}}$.\label{fig4}}
\end{figure}

Since ADI does not achieves a $\sqrt{n}$ noise attenuation gain at all separations, the bottom panel of Fig.~\ref{fig3} and Eq.~\ref{eqR} can be used to estimate the residual speckle decorrelation time that is currently limiting ADI performances. Fig.~\ref{fig4} upper panel shows the estimated decorrelation timescale as a function of angular separation. Typical decorrelation times are of the order 1-3 minutes and are generally smaller at larger angular separations. Normalizing these curves by the time $\rm{MAX} (\tau_{\rm{FWHM}}, t_{\rm{exp}})$ (see Fig.~\ref{fig4} bottom panel), it can be deduced that, for separations less than 3$^{\prime \prime}$, since $\tau_{\rm{dcorr}}$ is less than $\tau_{\rm{FWHM}}$, ADI differences are limited by residual speckles having $\tau_{\rm{speck}} < \tau_{\rm{FWHM}}$. These residual speckles evolve faster than the time needed to obtain an ADI reference PSF and cannot be subtracted by ADI. For HD97334B and HD18803, for separations greater than 3$^{\prime \prime}$, ADI is limited by the FOV rotation or the noise is decorrelated (limited by $t_{\rm{exp}}$). For Vega, the decorrelation time increases until $\tau_{\rm{FWHM}} = t_{\rm{exp}}$ and then decreases. The fact that $\tau_{\rm{dcorr}}$ is significantly greater than $\tau_{\rm{FWHM}}$ when $\tau_{\rm{FWHM}} \ge t_{\rm{exp}}$ suggests that Vega ADI differences, at those separations, are limited by correlated noises that are bigger than one FWHM. This noise thus requires more than a FWHM FOV rotation to be decorrelated. The upper limit at $\sim 1.5$ FWHM is expected since a minimal FOV rotation of 1.5 FWHM ($\tau_{\rm{min}}$) has been chosen to build the ADI optimized reference PSF. The ADI optimized reference PSF subtraction is a temporal filter that guarantees that no noise can have a decorrelation time longer than the time needed for a $\sim 1.5$ FWHM FOV rotation.

\subsection{Contrast Performances}
\setcounter{footnote}{3}

\begin{figure}
\epsscale{1}
\plotone{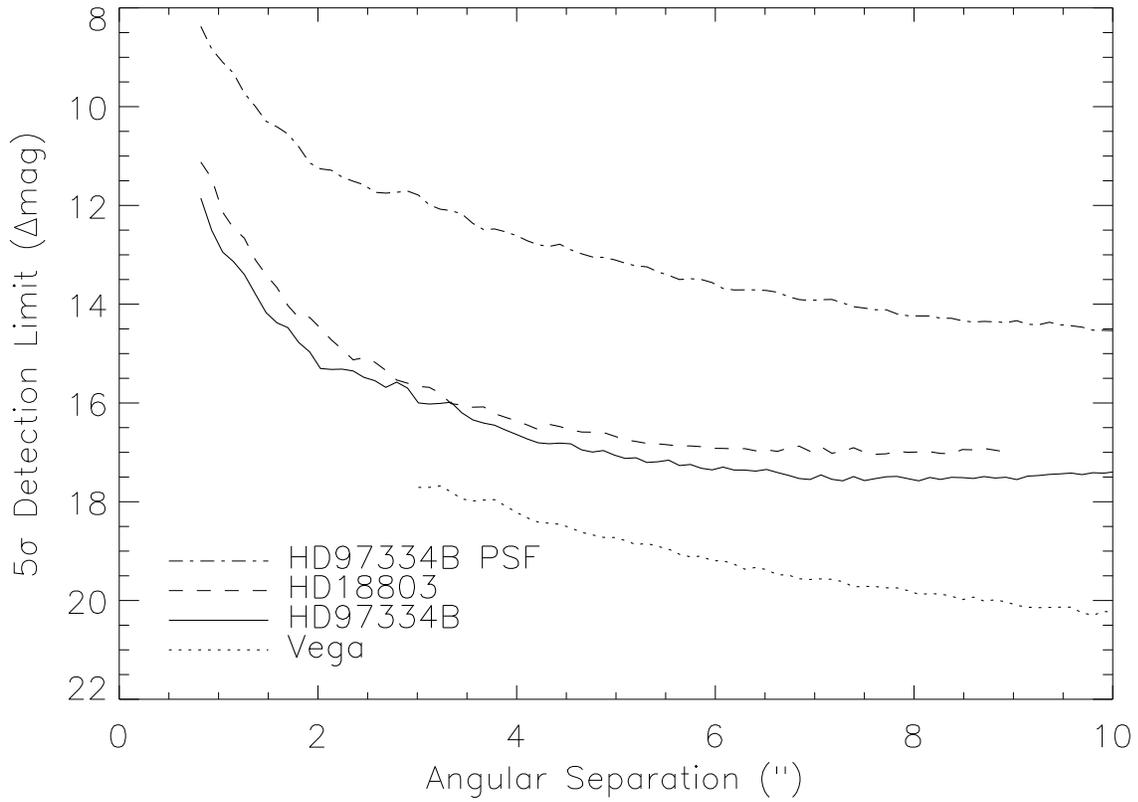}
\caption{Final ADI detection limits (5$\sigma$) as a function of angular separation for HD18803, HD97334B and Vega (August 26th). The initial detection limit for the HD97334B PSF (dot-dashed line) is also shown to illustrate the ADI noise attenuation performance. \label{fig5}}
\end{figure}

\begin{figure}
\epsscale{0.95}
\plotone{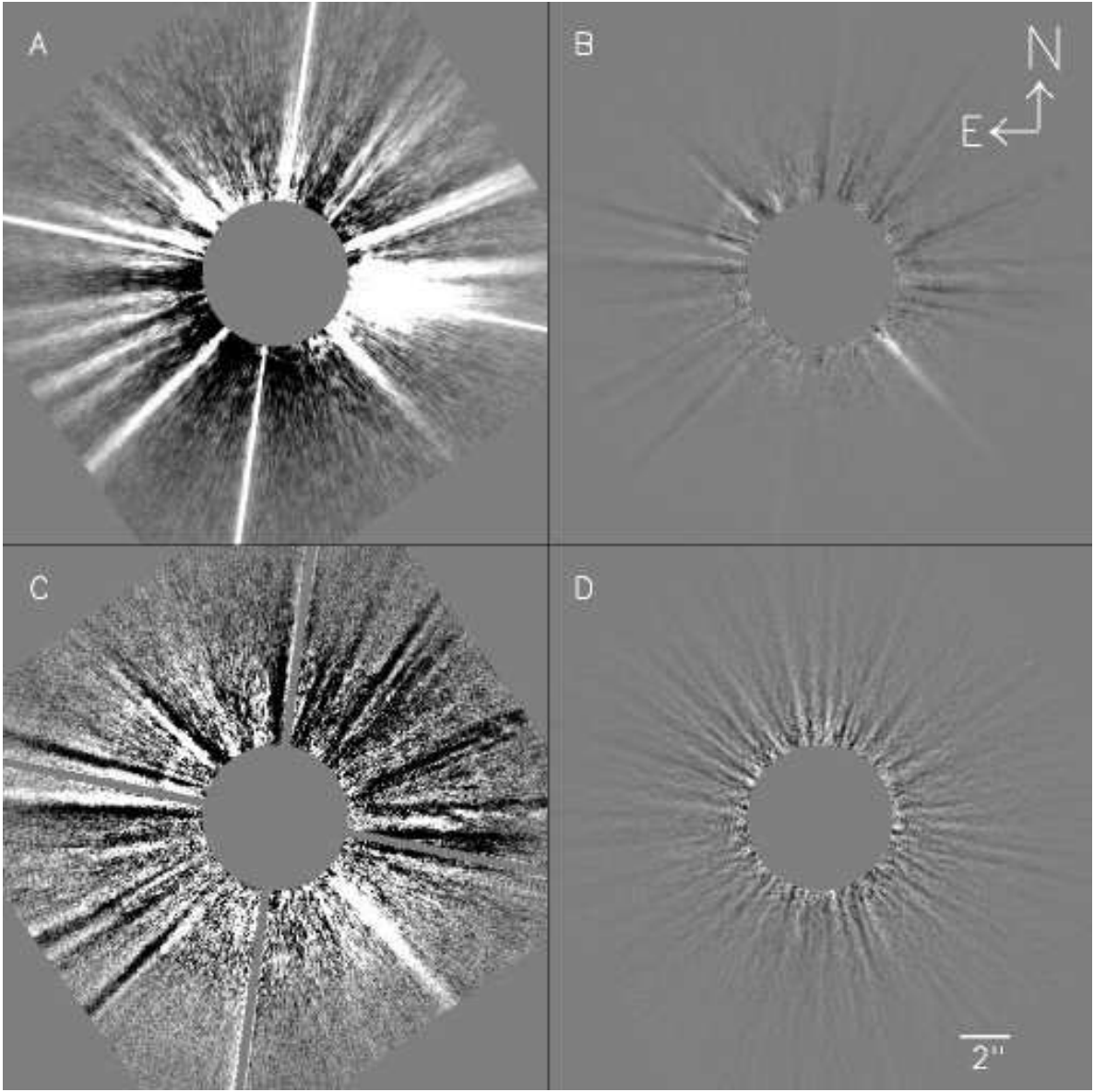}
\caption{Vega (August 26th) ADI data reduction. A: A single image after flat field
  normalization, bad pixel correction, distortion correction,
  registering and removal of an azimuthally symmetric profile. FOV is $22^{\prime \prime}\times 22^{\prime \prime}$ using a linear intensity range of $\pm 10^{-6}$ from the estimated PSF peak intensity. B: A single ADI difference image shown with the same intensity range. C: same as B with an intensity range 25 times smaller. D: The final combination of all ADI differences shown with the same intensity range as C. The central saturated 6$^{\prime \prime}$ diameter region as well as diffraction from the secondary mirror supports have been masked.\label{fig6}}
\end{figure}

Fig.~\ref{fig5} shows detection limits (5$\sigma$) in magnitude difference as a function of angular separation obtained with the ADI
technique for all three ADI targets. To produce this figure, data were reduced following the procedure explained in section~\ref{reduc}. No unsharp mask was used for the ADI reduction since multiple tests have shown that while this filter is effective for suppressing the low-frequency spatial noise, it does not improve candidate S/N and the photometry is slightly biased in the process. The detection limits are calculated using
the ratio of simulated companion peak intensities over the noise in the residual image as a function of angular separation. The flux normalized unsaturated PSF was then used to simulate smeared companion PSFs as a function of angular separation. The noise is calculated inside annuli of increasing diameter and width equal to 1~PSF FWHM. To account for the PSF smearing effect due to FOV rotation, the image and simulated companions were convolved with elliptical Gaussian of one FWHM in the radial direction and one FWHM plus a smearing term in the azimuthal direction, the smearing term ranging from zero at the center to typically two FWHM at 10$^{\prime \prime}$, depending on the rotation rate of the FOV. Finally, detection limits were corrected for the Altair estimated anisoplanatism following the Strehl $S$ equation found in the Gemini web page\footnote{http://www.gemini.edu/sciops/instruments/altair/altairIndex.html}

\begin{equation}
S(\theta) = S_0 e^{-\left(\frac{\theta}{12.5}\right)^2}
\end{equation}

\noindent where $\theta$ is expressed in arcsec. The detection limit obtained with the ADI technique on Vega at separations greater than $\sim$5$^{\prime \prime}$ is two orders of magnitude deeper than the Palomar $H$-band image \citep{metchev2003} and approximately a factor of ten deeper than the Keck $K$-band image of \citep{macintosh2003}. The current speckle attenuation (20-100) achieved with ADI is comparable or better to what is currently obtained (50) from one HST orbit using the roll subtraction technique. Although the ADI technique is inherently optimized for relatively large separations, the good Gemini PSF stability enables excellent performances at sub-arcsec separations. Indeed, the ADI contrast of $\Delta m = 11.1 - 11.9$ (5$\sigma$) at 0.8$^{\prime \prime}$ obtained on HD18803 and HD97334B equals the $\Delta m = 11.0$ (5$\sigma$) at 0.8$^{\prime \prime}$ obtained with the simultaneous differential imaging (SDI) VLT camera optimized for multi-wavelength speckle suppression \citep{biller2006}. This shows the potential of ADI to achieve high-contrast detection at sub-arcsec separations using a simple, yet efficient, observing technique with standard instruments. Fig.~\ref{fig6} illustrates the noise attenuation obtained for Vega (August 26th) using the ADI technique.

Mass limits corresponding to these observations, corrected for the filter use\footnote{The star-to-planet brigthness ratio is reduced by a factor $\sim$2.6 when using a 6.5\% bandpass methane filter instead of a broad band $H$, as derived from a theoretical spectrum of \citet{allard2001}.}, are estimated using evolutionary models of \citet{baraffe2003} assuming ages of 350, 45 and 85~Myr for Vega, HD18803 and HD97334B, respectively \citep{song2001,montes2001}. Both HD18803 and HD97334B achieve detection limits of 1-2 M$_{\rm{Jup}}$ at 3$^{\prime \prime}$ (60~AU for both targets), while $\sim$3~M$_{\rm{Jup}}$ is obtained for Vega at 8$^{\prime \prime}$ (63~AU). The ADI technique is thus well suited to survey jovian companions at intermediate separations (50-300~AU) orbiting young nearby stars.

\subsection{Comparison between ADI and Classical Imaging\label{adivscla}}

In the previous sections it was shown that the ADI technique can achieve
high contrast given a sufficiently long integration time and good
PSF stability.
To compare the performances of ADI and classical observations we analyze
the first 38 images of the HD97334B ADI sequence and the 38 images of the
HD1405 ``classical'' sequence. For this analysis, both data sets have
been reduced according to section~\ref{reduc}. Furthermore, an $8\times 8$ FWHM unsharp mask was applied to all images to remove the low spatial frequency quasi-static
noise. Then a $1\times 1$ FWHM median filter was applied to all
images to remove the bad/hot pixels. These steps are performed here only to bring the classical observations on even ground with ADI in order to study the evolution of the noise at spatial scales that most severely limit point source detections. For the HD97334B sequence, images differences were
obtained according to section~\ref{adialgo}; these differences were then rotated to align the FOV to that of the first image.

\begin{figure}
\epsscale{1}
\plotone{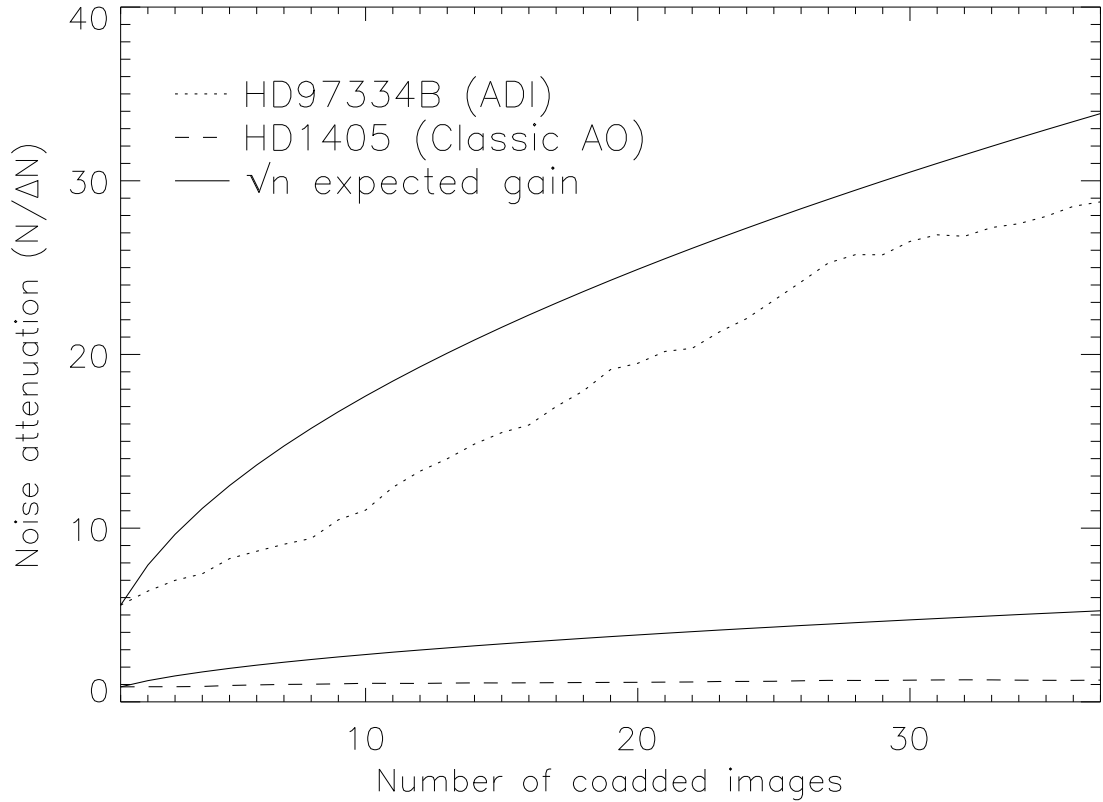}
\caption{Normalized noise attenuation from the median combination of an increasing
  number of images (differences in the ADI case) at 2$^{\prime \prime}$ separation. Each single image represents 30s exposure time. The dotted
  and dashed lines are respectively for the ADI reduction and the
  HD1405 without field rotation sequence. See section~\ref{adivscla} for more details.
\label{fig7}}
\end{figure}

An increasing number of images (differences for HD97334B) of both
sequences were median combined to study the noise attenuation as
a function of the total observing time at 2$^{\prime \prime}$, the results are presented in Fig.~\ref{fig7}. The ADI reduction technique achieves 30 times better speckle noise attenuation compared to classical AO observations in 30 minutes integration time. This figure also illustrates the power of ADI imaging in which noise attenuation, and thus companion S/N, increases nearly as the expected $\sqrt{n}$ while it saturates rapidly for normal imaging. 

\section{Discussion}

ADI is a general high-contrast imaging technique that can be applied to
any existing or upcoming large altitude/azimuth telescope. It is also 
flexible enough to be combined with a number of other high-contrast
imaging techniques.

ADI at small separations ($<$~1$^{\prime \prime}$) requires
long time intervals and thus may suffers from PSF variations. These variations come from variable seeing and slowly evolving quasi-static aberrations from the telescope and instrument optics. Unsaturated data inside 0.8$^{\prime \prime}$ are required to estimate performances. Extrapolation from our observations predicts that noise attenuation will be limited by speckles evolving faster than $\tau_{\rm{FWHM}}$ required for a 1 FWHM FOV rotation. Speckle noise attenuation of the order of a factor 10 should be feasible for one-hour sequences. The use of a
multi-wavelength instrument \citep{marois2000,doyon2004,lafreniere2004,marois2004,marois2005,biller2004} or an IFU
\citep{sparks2002} to acquire simultaneous images at multiple
wavelengths across the methane absorption bandhead at 1.6~$\mu $m through the
simultaneous spectral differential imaging (SSDI) technique \citep{racine1999,marois2000,biller2004,marois2004phd,marois2005}
could provide good short-lived and common-path speckle attenuations and increase detection
limits. It has been shown that SSDI instruments are ultimately limited by
non-common path aberrations, which are expected to be stable over long
periods of time as they come almost entirely from the instrument
itself \citep{marois2005}. Hence, the ADI technique nicely complements SSDI since it can be 
used to subtract the residuals caused by the non-common path aberrations.

Future high-contrast instrumentation for 8-10 m class or larger
telescopes based on high-order adaptive optics (AO) systems \citep{macintosh2004,mouillet2004} will most likely improve the stability of the PSF. Thus, if combined to such
instruments, ADI could prove even more successful. These new AO systems
will also provide the high Strehl ratios required to bring coronagraphy
at the forefront of high-constrast imaging. However, even a very good coronagraph cannot totally suppress
the light from uncorrected quasi-static wavefront errors and some level of quasi-static speckle
noise will inevitably be present in coronagraphic observations. ADI
would be a nice addition to coronagraphy as it could attenuate those
residual speckles.

Ideally, all the techniques mentioned above, ADI, SSDI, high-order AO
and coronagraphy, could be used together to form an extremely powerful
tool to detect exoplanets and brown dwarfs around stars.

\section{Conclusion}

The ADI observing technique was described and its performance using
Altair/NIRI at Gemini was presented. It was shown that faint companions can
be detected with better S/N when compared to classical
observing techniques for a wide range of declinations. The ADI technique produces a reference PSF from the same target imaging sequence, removing the need to move to a nearby star for PSF calibration or to acquire sky exposures (for $H$-band imaging). Since the reference PSF is built using images acquired minutes apart, the reference PSF shows a good quasi-static speckle correlation.

The stability of the PSF plays a crucial role in ADI as it not only
determines the speckle attenuation from the reference image subtraction
but it also determines the regime in which the noise is attenuated with
increasing observing time. It was reported that at Gemini with
Altair/NIRI using 30s exposures, the PSF evolves on timescales of $\sim 10-60$ minutes and
the attenuation by subtraction of a reference image reaches $\sim 2-6$
for short time intervals, achieving better speckle attenuation with better seeing conditions. The observations of HIP18859, for which a
filter change during the sequence reduced significantly the speckle
attenuation, underscore the importance of maintaining the optical setup
fixed during the sequence. It was shown that the gain in S/N with
increasing total observing time for separation greater than 2$^{\prime \prime}$ reaches more than 70\% of the optimal case,
indicating that the noise is mostly decorrelated between residual
images for these separations. Typical residual speckle decorrelation time is of the order of a few minutes. The speckle noise residuals decorrelate faster for object having faster FOV rotation. In all cases, ADI guarantees a larger gain with longer observation sequences. To our knowledge, this is the first time that such behavior is clearly demonstrated for an acquisition and reduction technique designed for speckle attenuation. The wall raised by quasi-static speckles that prevents a gain with longer integration time for standard observing techniques \citep{marois2003,marois2005,masciadri2005} can thus be removed by ADI. Comparison with a classical imaging technique shows that ADI achieves 30 times better speckle attenuation in 30 minutes integration time.

The noise attenuation obtained on Vega was 100, reaching a contrast of $\sim$20
magnitudes at 8$^{\prime \prime}$ separation (63~AU). Observations of the young stars HD18803 and HD97334B yielded
detection limits in difference of magnitude of 11.1-11.9 at 0.8$^{\prime
  \prime}$, similar to the SDI camera at VLT ($\Delta m$ of $\sim $11 at
0.8$^{\prime \prime}$), which is an optimized speckle suppression 
instrument. When combined to substellar models and estimated age for these stars, these observations show that ADI is well suited to search for jovian companions having a mass greater than 1-2 M$_{\rm{Jup}}$ 50-300~AU away from nearby young stars. Finally, ADI could easily and advantageously be combined with SSDI, high-order AO and coronagraphy to improve the detection limits of exoplanets and brown dwarfs at all separations.

\acknowledgments
This research has made use of the SIMBAD database, operated at CDS, Strasbourg, France. Authors would like to thank Ren\'{e} Racine for his comments on the manuscript, Fran\c{c}ois Rigaut and the Gemini observing staff for introducing a neutral density filter inside Altair that made Vega observations possible and Michael Fitzgerald, Paul Kalas, James Graham, Mike Liu, R\'{e}mi Soummer and \'{E}tienne Artigau for discussions about data reduction techniques and/or speckle statistic. The authors wish to recognize and acknowledge the very significant cultural role and reverence that the summit of Mauna Kea has always had within the indigenous Hawaiian community. We are most fortunate to have the opportunity to conduct observations from this mountain. This work is supported in part through grants from the Natural Sciences and Engineering Research Council, Canada and from the Fonds Qu\'{e}b\'{e}cois de la Recherche sur la Nature et les Technologies, Qu\'{e}bec. This research was also partially performed under the auspices of the US Department of Energy by the University of California, Lawrence Livermore National Laboratory under contract W-7405-ENG-48, and also supported in part by the National Science Foundation Science and Technology Center for Adaptive Optics, managed by the University of California at Santa Cruz under cooperative agreement AST 98-76783.

\clearpage
\begin{center}
\begin{table}
\begin{center}
\caption{Observations\label{tabobs}}
\begin{tabular}{ccccccc}\hline
Object & Date & Nb & Field rotation& $R_{\rm{min}}$ ($^{\prime \prime}$) & $\tau_{\rm{min}}$ (min) & Strehl\\
 & & images & (degree) & & at meridian \& 1$^{\prime \prime}$& \\
Vega & 08/26/04 & 225 & 99 & 0.15 & 8 & 0.24$^{\rm{a}}$\\
Vega & 09/01/04 & 177 & 69 & 0.1 & 8 & 0.12\\
HD18803 &  12/24/04 & 90 & 99 & 0.15 & 2 & 0.10\\
HD97334B & 04/18/05 & 90 & 54 & 0.3 & 6& 0.16 \\ \hline
\end{tabular}
\end{center}
{\footnotesize $^{\rm{a}}$ Short exposures for the August 26th run are saturated. Since the September 1st short exposures fill 55\% of the pixel electron well at the PSF peak intensity, it is approximated that the Strehl ratio for August 26th is at least 2 times higher than the one estimated for September 1st.}
\end{table}
\end{center}

\clearpage
\begin{center}
\begin{table}
\begin{center}
\caption{ADI Data Reduction Algorithm\label{tabadi}}
\begin{tabular}{lll}\hline
Reduction & Flat field normalization&\\ 
&Bad pixel correction&\\ 
&Distortion correction&\\ 
&Copy in bigger blank image&\\ \hline
Processing & Image registration&\\ 
&Radial profile subtraction&\\ 
&Diffraction spikes \& saturation masking&\\ \hline
ADI & Subtraction of the median of all images&\\
& For each 0.6$^{\prime \prime}$ wide annulus & Find 4 images with $\tau \ge 1.5\tau_{\rm{min}}$\\
&& Median combine these 4 images\\
&& Flux normalization\\
&& Subtract reference annulus\\
&Calculate parallactic angle&\\ 
&Rotate images&\\ \hline
Coaddition & Median combination of all differences&\\ 
&(optional) Convolution by Gaussian&\\ \hline
\end{tabular}
\end{center}
\end{table}
\end{center}
\clearpage

\clearpage

\clearpage

\clearpage

\clearpage

\clearpage

\clearpage

\clearpage

\end{document}